# Graphene-enhanced, internal-magnetic-field-generated Rabi oscillations in metal-coated Si-SiO$_2$ photoconductive detectors

**Venkatram Nalla[1,2, a)], Kai Zhang[3], Kian Ping Loh[3] and Wei Ji[2, a)]**

[1]*Centre for Disruptive Photonic Technologies, Nanyang Technological University, Singapore 637371*

[2]*Department of Physics, National University of Singapore, Singapore 117542*

[3]*Department of Chemistry, National University of Singapore, Singapore 117543*

a) Author to whom correspondence should be addressed. Electronic mail: vnalla@ntu.edu.sg, phyjiwei@nus.edu.sg.

**Abstract:**

We report our demonstration of Rabi oscillations in Si-SiO$_2$-Al photoconductive devices with nanosecond laser pulses of a few nJ at room temperature without external magnetic fields. Zeeman splitting of spin quantum states of dopants in silicon is achieved with internal magnetic fields produced by the Al film under excitation of laser pulses. Rabi oscillation frequency is 15 MHz and 25 MHz when photocurrent direction is perpendicular and parallel, respectively, to the propagation direction of linearly-polarized, 532-nm, 7-ns laser pulses. Insertion of graphene buffer layer between Al and SiO$_2$ provides a three-fold enhancement in Rabi oscillation amplitude. This simple-structured, low-cost device operated at room temperature should open a new avenue for future spin-based electronics and optoelectronics.

Electron spin is an intrinsic form of angular momentum carried by charge carriers (electrons or holes), which may give rise to many interesting electrical and optical properties of a material. Spin-related phenomena, such as spin coherence, spin mixing and spin transport, have been proposed to be utilized in next-generation devices in data processing, communications and storage. Realization and manipulation of Rabi oscillations, one of the spin coherent phenomena, have recently attracted attention as they were reported to be the first step in the quest for quantum computation.[1-3] Rabi oscillations manifested themselves in the photocurrents of semiconductor devices whereby various quantum dots made of compounds like GaAs,[3,4] InGaAs,[5] GaAs/AlGaAs,[6] or InAs[7,8] were embedded. Spin Rabi flopping was also observed in the photocurrent of a polymer light-emitting diode.[9] Furthermore, single vacancy defect centers could also lead to Rabi oscillations in the fluorescence intensity.[10] Very recently, ultrafast Rabi oscillations between excitons and plasmons were reported for metal nanostructures with J-aggregates.[11]

Silicon is the widely used semiconductor in electronic, optoelectronic and photonic devices; and it is the mostly available material on earth. Realization and manipulation of Rabi oscillations in silicon-based devices would have direct impact onto present-day information technology. A number of research groups studied spin coherence phenomena in silicon-based photoconductive devices as discussed in the following. Phosphorus donors ($^{31}$P) in silicon offer spin quantum states.[12] When exposed to light, spin quantum states of $^{31}$P give rise to Rabi oscillations in the photocurrent, due to coherent manipulation of spin-dependent charge-carrier recombination between spin quantum states of $^{31}$P and paramagnetic localized states of $P_{b0}$ centers at the Si/SiO$_2$ interface.[13] Coherent manipulation of an individual electron spin qubit bound to a phosphorus donor atom in natural silicon has also been demonstrated as single-atom electron spin qubit.[2]

It should be pointed out that all the above studies on Rabi oscillations in silicon-based devices were carried out with externally applied magnetic fields for the attainment of Zeeman splitting.[2,13-16] Microwave fields were also utilized to generate Rabi oscillations in silicon devices as well.[17] Here we report our realization of Rabi oscillations in metal-coated Si-SiO$_2$ photoconductive devices without external magnetic fields or microwave fields. We also demonstrate that such Rabi oscillations can be enhanced by insertion of graphene layer between metal film and SiO$_2$ layer. The realization of Rabi oscillations in the above structure can be conceptualized as follows.

As illustrated in Figure 1, when irradiated by electromagnetic waves, nano- or micro-scale, irregularly-shaped cracks, normally present in a thermally vapor deposited metal film (such as Au, Ag, Ni, Al, etc...) on a dielectric substrate can generate stronger magnetic fields (about three orders as compared to the magnetic field of the incident wave) at the interface, predicted by theoretical calculations and supported by experimental observation.[18-22] It is also estimated that if metal ring cavities exist in the metal film, magnetic fields of up to 400 mT can be generated with an optical excitation.[23] On the other hand, similar to phosphorus dopants in silicon, boron dopants also exhibit significant Zeeman splitting at magnetic fields as low as ~100 mT (or, ~10 KG).[24,25] High magnato-resistance resulting from internal magnetic fields found in Al-coated Si-SiO$_2$ structures give a consistent account.[26] By a combination of these two factors, we believe that Rabi oscillations should occur in metal-coated Si-SiO$_2$ structure without external magnetic fields, as shown in Figure 1. When light is shone to the metal film, magnetic fields are generated due to the existence of nano- or micro-sized cracks. As a consequence, Zeeman splitting of spin quantum states of donors (such as $^{10}B$)[27] in silicon gives rise to Rabi oscillations in the measured photocurrents.

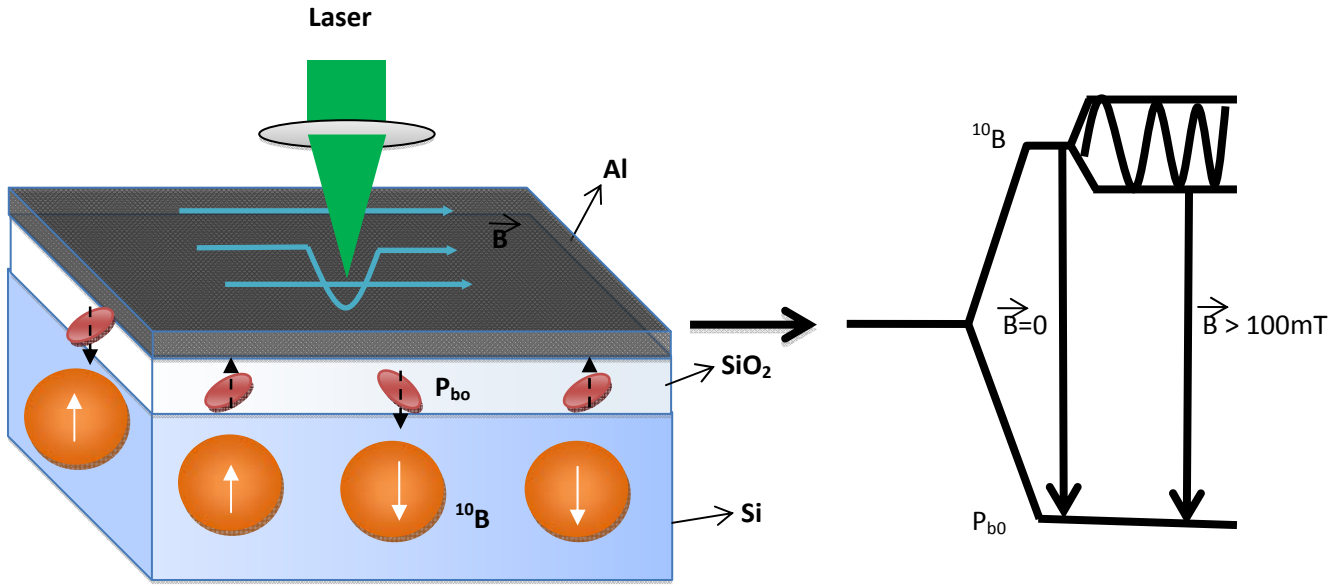

Figure 1: Schematic of internal magnetic fields generated from the metal film and Zeeman splitting of dopant spin quantum states in silicon in metal-coated Si-SiO$_2$ structure.

To demonstrate the above concept, we designed and fabricated two simple photoconductive detectors with different structures by utilization of commercially available, low-cost silicon and aluminium. Acted as both electrodes and magnetic-field generators, aluminium (Aldrich) was coated on Si-SiO$_2$ wafers (NOVA Electronic Materials, <100>, 10 ohm-cm, 500-μm-thick Si and ~100 nm-thick SiO$_2$) with thermal vapour deposition technique. In Device 1 shown in Figure 2(a), two Al electrodes with a thickness of 100 ± 20 nm and a separation of 1 mm were deposited on the top surface of the Si-SiO$_2$ wafer. In Device 2, two Al films with the same thickness (~ 100 nm) were coated on both top and bottom surfaces of the Si-SiO$_2$ wafer, as shown in Figure 2(b).

Transient photocurrent measurements were performed on the two devices in short circuit by connecting an oscilloscope directly to the Al electrodes. Note that neither external electrical voltages nor external magnetic fields were intentionally applied to the devices. Schottky barriers at the Si- SiO$_2$ interface separate electron-hole pairs generated by absorption of incoming laser beam. As the SiO$_2$ layers were so thin that photo-generated charge carriers in silicon can efficiently tunnel through the SiO$_2$ barrier to the Al film. The laser beam was provided by the second harmonic generation of a Q-swictched Nd:YAG laser (Spectra Physics, Quanta Ray) with 532-nm wavelength, 7-ns duration, and 10-Hz repetition rate. The linearly polarized laser beam was focused and normally incident onto the device with a spot size of ~100 μm using a 20-cm, focal-length lens. Upon on laser pulsed radiation, the temporal photocurrent responses were recorded with a digital oscilloscope (Tektronix TDS 380, 400 MHz bandwidth); and the positive and negative terminals (or, anode and cathode) of the oscilloscope were directly connected to the two Al electrodes, respectively. All the measurements were carried out at room temperature and ambient condition.[28,29]

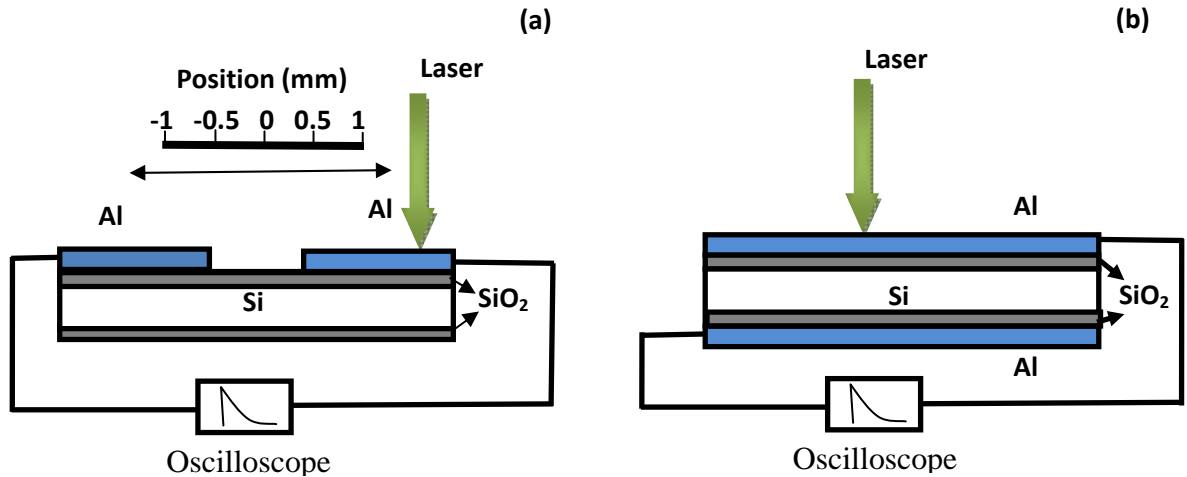

Figure 2. Schematic of Si-SiO$_2$-Al photoconductive devices: (a) two Al electrodes on the top surface of SiO$_2$-Si-SiO$_2$ (Device 1) and (b) one Al electrode on the top surface and the other is on the bottom surface of SiO$_2$-Si-SiO$_2$ (Device 2). When the devices are exposed to laser beams, photocurrents were measured directly between the two Al electrodes by an oscilloscope symbolized by the box without external electrical voltage and external magnetic fields.

Figure 3(a) shows the measured transient photocurrents by directing the laser beam at different position with respect to the center of Device 1. Temporal profiles of the transient photocurrents show clearly exponential decay curves superimposed with oscillatory behavior when laser excitation is on the Al electrode. Figure 3(a) demonstrates the position-dependent Rabi oscillations in Device 1. Negative and positive photocurrent decay curves with oscillatory behavior are produced if the laser beam is incident on the cathode and anode, respectively; and a maximum magnitude of Rabi oscillation is found around ±1.2 mm away from the center. This position-dependent Rabi oscillation clearly demonstrates that the magnetic field generated from the Al film is responsible for the manifestation of Rabi oscillations in the photocurrent. It is consistent with the observations of high internal magnetic fields (10 - 300 mT) by other research groups in thermally vapor deposited metal films wherein micro/nano-sized irregular-shaped cracks are present, and utilized as *q*-bits for quantum computing.[30,31] Because of the presence of internal magnetic fields, Zeeman splitting occurs between the spin quantum states of $^{10}$B in silicon, leading to spin-dependent charge-carrier recombination between spin quantum states of $^{10}$B and paramagnetic localized states of $P_{b0}$ centers at the Si-SiO$_2$ interface.

If the laser beam is shifted to the center between the Al electrodes in Device 1, oscillatory behavior disappears and only decay curves are observed. In the center of the device where magnetic fields are close to zero and hence, no Rabi oscillations happen; and the net positive photocurrent is mainly due to the difference between the mobility of photo-generated electrons and holes, similar to other planar photoconductive devices.[28]

Fourier transformed photocurrent oscillations in Devices 1 and 2 are shown in Figure 4(a) and 4(b), respectively. From these figures, we find that the Rabi oscillation frequency is ~ 15 MHz in Device 1 and ~ 25 MHz in Device 2; and both are independent of the laser excitation energy, see Figure 4(c). Since the Rabi oscillation frequency linearly depends on the magnetic field,[14,17] magnetic fields should be higher in Device 2 than Device 1. This is consistent with the theory,[20] which predicts that magnetic fields generated from the metal film are considerably greater in the beam propagation direction (utilized in Device 2) than the perpendicular direction (utilized in Device 1). Both theory and experiment[18,32] point out that the magnetic field depends on incident light intensity (or square of incident electromagnetic fields) and is saturated in the regime of high light intensity. In our experiment, we measured photocurrents in the saturation regime and it is why the Rabi oscillation frequencies are independent of incident light intensity. We were unable to measure photocurrents at lower light intensities, due to the poorer responsivity (~10 mA/W) of our devices, as compared to commercially available silicon photoconductive detectors (responsivity > 100 mA/W) whereby an external bias provides efficient separation of photo-generated charge carriers.

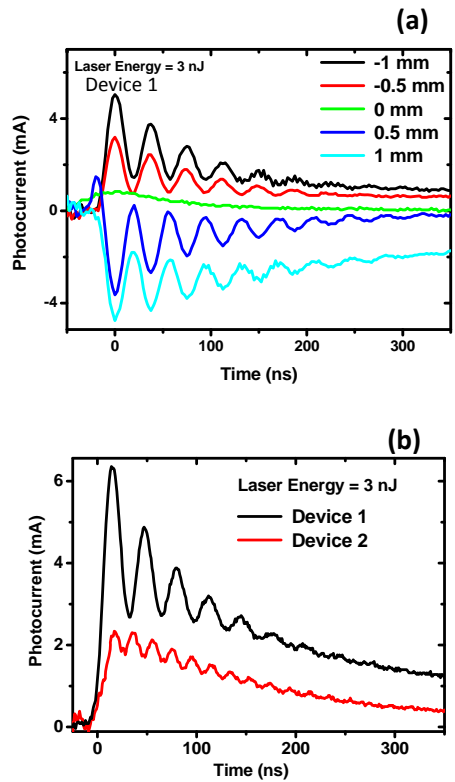

Figure 3. (a) Laser excitation position dependence of transient photocurrent decays with Rabi oscillations in Device 1 and (b) comparison of maximum transient photocurrent decays with Rabi oscillations between Device 1 and Device 2.

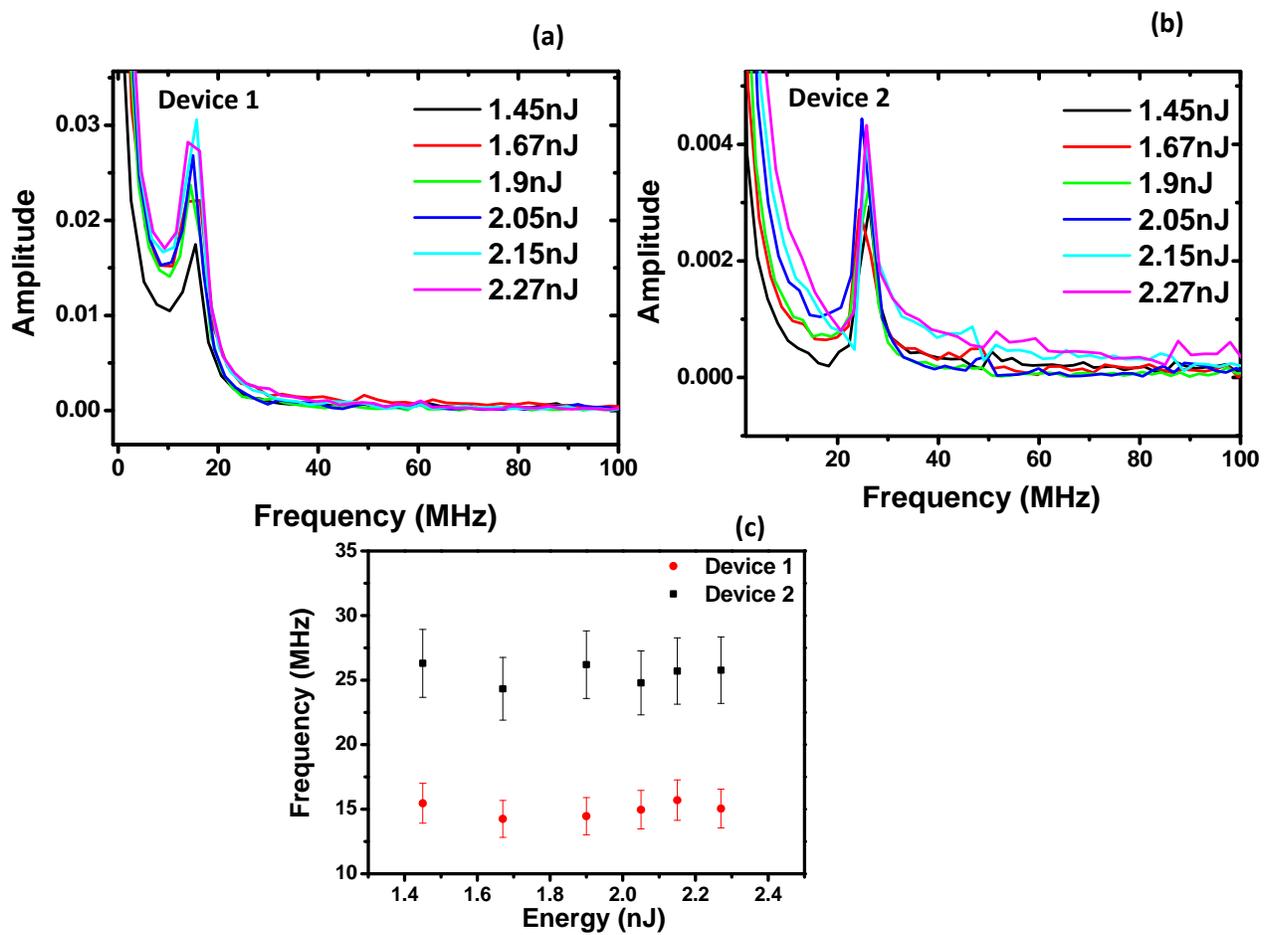

Figure 4. Fourier-transformed photocurrents are plotted as a function of frequency in (a) Device 1 and (b) Device 2. In (c), Rabi oscillation frequency is plotted against laser excitation energy.

Graphene is known to possess high electron and hole conductivity. Many research groups found that adding graphene buffer layers can enhance both solar cell efficiency and photodiode gain, resulting from efficient collection of charge carriers.[33-36] We believe that adding graphene buffer layer between Al and $SiO_2$ should also enhance the collection of photo-generated charge carriers as well. As such, we made an extra device (Device 3) that has the same structure as Device 1 except for monolayer graphene being placed between Al and $SiO_2$ layer. Following the previously reported procedures,[37] monolayer graphene was grown by low-pressure chemical vapor deposition (CVD) and transferred onto the top surface of the Si-$SiO_2$ wafer prior to the coating of Al electrodes. Figure 5 displays the Fourier-transformed photocurrents of Devices 1 and 3. It supports that the amplitude of Rabi oscillation is as three times in Device 3 as in Device 1.

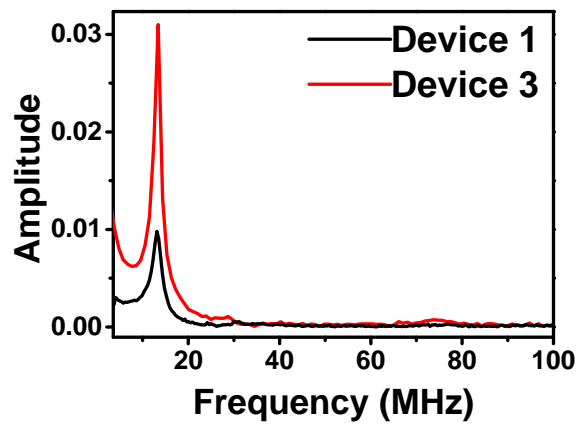

Figure 5. Rabi oscillation amplitudes as function of frequency in Devices 1 and 3.

In summary, we have demonstrated Rabi oscillations in Si-SiO$_2$-Al photoconductive devices with nanosecond laser pulses of a few nJ at room temperature without externally magnetic fields. Zeeman splitting of spin quantum states of dopants in silicon is achieved with internal magnetic fields produced by the Al film under excitation of laser pulses. Rabi oscillation frequency is 15 MHz and 25 MHz when photocurrent direction is perpendicular and parallel, respectively, to the propagation direction of linearly-polarized, 532-nm, 7-ns laser pulses. Insertion of graphene buffer layer between Al and SiO$_2$ provides a three-fold enhancement in Rabi oscillation amplitude. This simple-structured, low-cost device operated at room temperature should open a new avenue for future spin-based electronics and optoelectronics.

The authors would like to acknowledge the Faculty of Science, National University of Singapore, Centre for Disruptive Photonic Technologies, Nanyang Technological University and Ministry of Education (MOE2011-T3-1-005) for financial support.